\documentclass[12pt,preprint]{aastex}

\shorttitle{Stellar stream candidates in the solar neighborhood}
\shortauthors{Liang et al.}

\begin{document}
\title{Stellar Stream Candidates in the Solar Neighborhood Found in the LAMOST DR3 and TGAS}
\author{X. L. Liang\altaffilmark{1,2}, J. K. Zhao\altaffilmark{1},  T. D. Oswalt\altaffilmark{3}, Y. Q. Chen\altaffilmark{1,2}, L. Zhang\altaffilmark{1}, G. Zhao\altaffilmark{1,2}}

\altaffiltext{1}{Key Laboratory of Optical Astronomy, National Astronomical Observatories, Chinese Academy of Sciences,
Beijing 100012, China. zjk@nao.cas.cn}
\altaffiltext{2}{School of Astronomy and Space Science, University of Chinese Academy of Sciences, Beijing 100049, China}
\altaffiltext{3}{Embry-Riddle Aeronautical University 600 S. Clyde Morris Blvd. Daytona Beach FL, USA, 32114. oswaltt1@erau.edu}

%\affil{1.Key Laboratory of Optical Astronomy, National Astronomical Observatories, Chinese Academy of Sciences, Beijing 100012,China.}
%\affil{2.School of Astronomy and Space Science, University of Chinese Academy of Sciences, Beijing 100049, China}
%\email{aastex-help@aas.org}

\begin{abstract}

We have cross-matched the LAMOST DR3 with the Gaia DR1 TGAS catalogs and obtained a sample of 166,827 stars with reliable kinematics. A technique based on the wavelet transform was applied to detect significant overdensities in velocity space among five subsamples divided by spatial position. In total, 16 significant overdensities of stars with very similar kinematics were identified. Among these, four are new stream candidates and the rest are previously known groups. Both the U-V velocity and metallicity distributions of the local sample show a clear gap between the Hercules structure and the Hyades-Pleiades structure. The U-V positions of these peaks shift with the spatial position. Following a description of our analysis, we speculate on possible origins of our stream candidates.

\end{abstract}


\section{Introduction}

The formation and evolution of galaxies is a topic at the forefront of astrophysical research. The solar neighborhood provides exquisite clues to the various processes that influenced the formation and evolution of the Milky Way. These kinematical and chemical processes can be used to constrain Galactic evolution models. Some of the most intriguing features in the solar neighborhood velocity field are moving groups (streams), which are manifested as overdensities in velocity space. For a long time it was believed that the structures were associated with disrupted stellar clusters \citep{kap05,egg65,sku97}. However, \citet{deh98} pointed out that most moving groups in the solar neighbourhood could be caused by orbital resonances. Observations \citep{fam05,fam07,ben07,fam08} indicate that stellar cluster disruptions are not responsible for most of the best well-known, such as the Sirius, Hyades, Pleiades and Hercules streams. Consequently, theoretical work tends to explain moving groups in terms of the dynamical effects of the bar and spiral arms of the Milky Way. The first theoretical arguments in favor of a different dynamic origin for moving groups were put forward by \citet{may72}. \citet{wei94} showed that the Galactic bar can lead to distinctive stellar kinematics near the Outer Lindblad resonance (OLR). Recently, because the Hercules structure does seem to have arisen from the resonant effects of the Galactic bar, the hypothesis of the dynamical resonant origin for these streams has become more popular \citep{kal91,deh00,fux01}.

Many theoretical models suggest that bar or spiral patterns can induce structures in the velocity space \citep{fux01,qui05,min06,ant09,gar10,lep11}. \citet{deh00} concluded that a group of unstable chaotic orbits produced the valley between the Hercules structure and the majority of stars in the U-V distribution. \citet{cha07} argued that effects of the bar and spiral arms must both be included to accurately reproduce the velocity structures seen in the solar neighborhood \citep{qui03,qui05,min06,cha08,ant09,min10,minc10,min11}. However It is difficult to replicate multiple spiral patterns by including the effects of both spiral and bar perturbations in numerical simulations. It is even more difficult to estimate the parameters of the arms by eye. To investigate the origin of these velocity structures, Antoja et al.\citeyearpar{hel12,ant14} analysed the velocity distribution outside the solar neighborhood for the first time using the RAdial Velocity Experiment (RAVE, \citealt{ste06}) data. They found several moving groups as large scale features and the Hercules structure shrinks as galactocentric distance increases. Recently,  \citet{mon16} used data from LAMOST and Gaia and found that Hercules disappears beyond R = 8.6 kpc.

The LAMOST (Large Sky Area Multi-Object Fiber Spectroscopic Telescope) \citep{cui12,zhao06,zhao12} is a special reflecting Schmidt telescope with an effective aperture of 3.6 - 4.9 m. LAMOST has a large field of view that can observe up
to 4000 objects simultaneously. The resolving power of LAMOST spectra is R $\backsim$ 2000 spanning $3700 \rm\AA  \thicksim 9000 \rm\AA $. The ESA (European Space Agency) Gaia mission \citep{gil12} is currently obtaining highly accurate parallaxes and proper motions of over one billion sources brighter than G $\backsimeq$ 20.7. The first data release (Gaia DR1) became available on 14 September 2016 \citep{gai16}. The primary astrometric data set in this release lists the positions, parallaxes, and proper motions of 2,057,050 stars that are in the Tycho-2 catalogue. This data set is called the Tycho-Gaia astrometric solution (TGAS).

The objective of our project is to find new overdensities and better understand the velocity structures in the solar neighborhood by combining LAMOST and Gaia databases. In Section 2, we describe the data used for stellar streams detection. Section 3 characterizes in detail the structures found in the U-V velocity plane. Finally, the main conclusions of our work and expectations for the future are summarized in Section 4.

\section{DATA}

We cross-referenced the LAMOST DR3 data and Gaia DR1 TGAS data and found 166,827 stars in common with $\log g \geq 4$. Then, five subsamples shown in Figure 1 were selected following the procedure of \citet{ant12}. We used the Galactic Cartesian coordinate system with the Sun at $X = -8.2$ kpc, $Y = 0$ kpc and $Z = 0$ kpc. First, the local sample (LS, S1) was defined as those stars with $|Z| \leq 300$ pc and in-plane distances from the Sun $dcos(b) \leq 200$ pc (lime green points) yielding 27290 stars in the LS. We then selected stars in four other rectangular regions, defined as:
    S2: -8.5 kpc $\leq$ X $<$ -7.9 kpc, -0.9 kpc $\leq$ Y $<$ -0.3 kpc, $|Z| \leq 0.3$ kpc (blue points);
    S3: -8.9 kpc $\leq X <$ -8.5 kpc, -0.3 kpc $\leq Y <$ 0.3 kpc, $|Z| \leq$ 0.3 kpc (aquamarine points);
    S4: -9.5 kpc $\leq X <$ -8.9 kpc, -0.4 kpc $\leq Y <0.4$ kpc, $ |Z| \leq 0.4$ kpc (yellow points); and
    S5: -8.5 kpc $\leq X < $-7.9 kpc, -0.3 kpc $\leq Y < 0.3$ kpc, 0.3 kpc $\leq |Z| \leq 0.9$ kpc (red points).
These regions probe different directions from the sun, providing information on how potential streams are oriented in space. The size and position of these regions were chosen by considering space distribution of the total sample and the position of the Sun. We tried to include more stars in each sample and avoid taking up too much space for each sample in the meantime. The velocity components in heliocentric coordinates were transformed into components U, V, and W in galactic coordinates. We took the peculiar solar motion as ($U_{\bigodot}, V_{\bigodot}, W_{\bigodot}$) = (-10.0, 5.2, 7.2) km $\rm s^{-1}$ \citep{deh98}. To calculate the errors of space motions, the uncertainties in parallaxes were adopted as $\omega\pm\sigma\omega$(random)$\pm0.3$ mas(systematic). We included the covariance of astrometry as well. The errors of UVW were calculated as  follows:
\begin{eqnarray}
  \left(\begin{array}{c}
 eU^2 \\
 eV^2 \\
  eW^2 \end{array} \right)
   = C \left( \begin{array}{c} evr^2 \\
   k^2*(\frac{\sigma_{\mu_\alpha}^2}{\pi^2}+\frac{\sigma_\pi^2*\mu_\alpha^2}{\pi^4})-2k^2*\rho_{\pi\mu_\alpha}*\frac{\mu_\alpha*\sigma_\pi*\sigma_{\mu_\alpha}}{\pi^3}  \\
   k^2*(\frac{\sigma_{\mu_\delta}^2}{\pi^2}+\frac{\sigma_\pi^2*\mu_\delta^2}{\pi^4})-2k^2*\rho_{\pi\mu_\delta}*\frac{\mu_\delta*\sigma_\pi*\sigma_{\mu_\delta}}{\pi^3} \end{array}
\right)+
 \left( \begin{array}{c}
 H_{12}H_{13} \\
 H_{22}H_{23} \\
 H_{32}H_{33} \end{array}
\right)*
\nonumber
\end{eqnarray}
\begin{eqnarray} 2k^2*(\frac{\rho_{\mu_\alpha\mu_\delta}*\sigma_{\mu_\alpha}*\sigma_{\mu_\delta}}{\pi^2}-\frac{\rho_{\pi\mu_\alpha}*\sigma_\pi*\sigma_{\mu_\alpha}*\mu_\delta}{\pi^3}
 -\frac{\rho_{\pi\mu_\delta}*\sigma_\pi*\sigma_{\mu_\delta}*\mu_\alpha}{\pi^3}+\frac{\mu_\alpha*\mu_\delta*\sigma_\pi^2}{\pi^4})
\textit{. } \nonumber
\end{eqnarray}

  For the above equation, \textit{evr} is the error of radial velocity, \textit{k} is the equivalent of 1 A.U./yr in km/s; $\mu_\alpha$ and $\sigma_{\mu_\alpha}$ are the components of proper motion in right ascension and their uncertainties; $\mu_\delta$ and $\sigma_{\mu_\delta}$ are the proper motion components in declination and their uncertainties; $\pi$ is parallax and $\sigma_\pi$ is the uncertainty in parallax; $\rho_{x_i,x_j}$ is the correlation coefficient between $x_i$ and $x_j$; \textit{H} is coordinate transformation matrix and $H_{ij}$ is the element of \textit{H} at ith row and jth column. The elements of the matrix \textit{C} are the squares of the corresponding elements of matrix \textit{H}. In addition, we ignored the errors of \textit{RA} and \textit{DEC} which are essentially negligible. Stars with any velocity error larger than 30 km $\rm s^{-1}$ were rejected.

To detect velocity structures in these sub-samples, a technique based on the wavelet transform (WT) \citep{zhao09} was used to search for overdensities in the velocity distribution. Using a bootstrap method, 100 simulated samples were generated, which selects the same number stars as the observed sample by random sampling with replacement. Then a WT was applied to each of the simulated samples.

\section{STRUCTURES IN THE SOLAR VICINITY}

\subsection{Velocity structures in LS} \label{bozomath}

 In our analysis we focused mainly on the U and V space velocity components, because moving groups have almost indistinguishable mean W velocity components \citep{deh98,sea07}. We chose 6 km $\rm s^{-1}$ as the scale of the mother wavelet used in the computation of the WT. We tried many different scales, and structures of this scale were the most detectable. Figure 2 shows the WT coefficients as a contour distribution of stars in the LS. Peaks are highlighted in blue while valleys are highlighted in red. This velocity distribution exhibits a number of clear overdensities. Those ten peaks marked by asterisks show up more than 90 times and six peaks marked by plus signs show up more than 70 times in the 100 simulated samples. These peaks are numbered from maximum to minimum WT coefficient and are listed in Table 1. The columns of Table 1 give, respectively: order in WT coefficient, name of the group in previous literature (if any), U  velocity and V velocity of peaks (Unless noted otherwise, all peak numbers mention below refer to those listed in Table 1.), approximate number of stars and the mean and variance of metallicity from Gaussian fitting. The member stars of each overdensity represented by a given peak are within 8 km/s from each peak in velocity space. Next we took metallicity as the independent variable and star counts in each unit metallicity as the dependent variable, performed a Gaussian Fitting, and computed the mean and variance. The last column in Table 1 shows the equivalent groups from RAVE \citep{ant12}. The peaks marked 9, 12, 13, 14 are four new stream candidates. We identify here positions of well-known streams agreed well with those from RAVE (separation between maxima $ < 3$ km $\rm s^{-1}$ ). However, there is one exception, the Coma Berenices stream. We did not find a peak around (U, V) = (-7, -6) km $\rm s^{-1}$. In addition, peak 15 cannot be declared as a new stellar stream, because \citet{ant12} marked it in the GCS \uppercase\expandafter{\romannumeral3}, although they did not claim it as a new stream or name it. Figure 2 contains two obvious regions: an inner part and an outer part. The inner part consists of five streams: Sirius stream (peak 2), Hyades stream (peak 0), Pleiades stream (peak 1), Dehnen98 stream (peak 5) and Wolf630 stream (peak 3); the other peaks belong to the outer part. We detected the famous Hercules structure at V = -50 km $\rm s^{-1}$, which includes the $\epsilon$ Ind stream (peak 8), the Hercules streams (peaks 6,4) and NEW1 (peak 9).

\subsection{Metallicity distributions} \label{bozomath}

Figure 3 shows the metallicity distribution of the LS. We divided the whole region into square bins with size 3 km $\rm s^{-1}$, and used color to represent the mean metallicity of each bin. The most prominent feature is Hyades-Pleiades structure, which includes the Hyades stream (peak 0) and Pleiades stream (peak 1). It exhibits a very metal-rich and consistent distribution; it is well known that the Hyades cluster has a high metallicity ([Fe/H]$\thicksim$0.13, \citealt{hei13}). We speculate that some stars of the Hyades stream might be remnants of the Hyades cluster, while others may be the dynamical effect of non-axisymmetric components \citep{fam08}. Though not in sharp contrast, there is a valley with lower metallicity between the horizontal V = -50 km $\rm s^{-1}$ structure and center. To better present the gap, we picked up a window in the metallicity distribution. We divided the window into parallelogram bins, and the edge parallel to the V axis of each bin equals 1 km $\rm s^{-1}$. The right panel of Figure 4 shows that the mean metallicity of each bin distributes along with the V direction. In Figure 3, the Hercules \uppercase\expandafter{\romannumeral1} (peak 6) and Hercules \uppercase\expandafter{\romannumeral2} (peak 4) have different metallicities, and peak 4 is definitely different from peak 12 (NEW2). That the Hercules stream has a wide metallicity range (Table 1) is in accord with the purported effect of Galactic bar resonances \citep{kal91,deh00,fux01}. The upper left side of this plot contains stars moving faster than the LSR. In this region the kinematics suggest peak 13 and peak 14 may be part of the thin disk, although their metallicities are quite low. The Sirius stream (peak 2) is a little more metal-poor than its surroundings and has a very small metallicity scatter. This is similar to the results of \citet{kle08}, and this supports the notion that the Sirius stream (or at least a part of it) are remnants evaporated from a cluster. According to \citet{sil07}, HR1614 (peak 11) may be dispersed remnants of star-forming events. New1 (peak 9) is close to peak 11 in both kinematics and metallicity; part of its member stars may be remnants of star-forming events, too. New2 (peak 12) has a large metallicity scatter. It is possible to be induced by other dynamical n:1 resonance of the bar. The $\gamma$ Leo stream (peaks 7, 10) is more metal-poor than our Sun, and it is rapidly moving toward the center of the Galaxy. Considering there is a radial metallicity gradient in the Galaxy, we conjecture this structure has drifted inwards from beyond the Solar Galactocentric radius.

\subsection{STRUCTURES BEYOND THE LOCAL VOLUME} \label{bozomath}

Due to observational constraints, the farther a region is from the Sun, the fewer stars it contains per unit volume. Thus the volumes of S4 and S5 are a little bigger than S2 or S3. There are 12693, 29001, 7036, 13689 stars in S2, S3, S4, S5, respectively. The four panels of Figure 5 show the velocity structures of S2, S3, S4, S5. Each of these four regions has a peak identified with the Coma Berenices stream. Peak 2 in panel (c) and Peak 7 in panel (d) coincide with the valley in Figure 2. From panel (b)-(c), we conclude that the Hercules structure diminishes as the distance increases, which is in accord with the results of \citet{deh99}. Several famous groups are visible in panels (a), (b), (d): Hyades, Pleiades, Sirius, Wolf 630 and $\epsilon$ Ind.  New4 can be seen in panel (b) and New3 can be seen in panel (d), but their relative positions change a little.  New1 is identifiable in panel (c) and (d), though its position changes a little, too. The overall shape of S4 is quite different from the LS. In panel (c), peaks marked by asterisks show up more than 80 times while peaks marked by plus signs only show up more than 50 times in the 100 times simulations. Our results suggest that some streams extend to great distances.

\section{ANALYSIS and CONCLUSION}

Non-axisymmetric Galactic potential models, especially those describing the spiral pattern and the Galactic bar, are often invoked to explain velocity structures in the solar neighborhood. For example, the dynamical effect of the Galactic bar is the most accepted explanation for the Hercules stream. As simulations by \citet{deh00} showed, there is no bimodality where the OLR is farther from the Galactic center than the Sun, and our findings are in agreement. It is hard to explain all the horizontal structures in the OLR as effects of the bar only, even when different resonant orbit families are invoked, although the 4:1 resonance of the bar is used to explain the Hercules structure. Spiral arms must play a role too. We compared our UV velocity distributions with a simulation (the right panel of Figure 2 from \citealt{hel12}), and found the profile of our distribution is similar to that of the 4:1 resonance which refers to the inner Lindblad resonance with radial oscillations whose ratio of orbital periods equal 4:1 \citep{fux01}. That suggests proximity to an arm strongly affects the velocity distributions of stars in the solar vicinity. We cannot rule out the possibility that those three peaks in the left-upper side of Figure 2 might originate from outside of the Milky Way (i.e. accretion events). They are metal-poor but their metallicity scatter is not small enough to suggest a disrupted cluster. However, if the velocity structures in Figure 2 are all induced by dynamical resonance of spiral arms, those three peaks may correspond to the structure at the left-upper side in last panel of column 2 from Figure 18 of \citet{ant11}. Also the $\gamma$ Leo stream (peaks 7, 10) and Hercules structure may correspond to the right-upper side and the lower side structure of that panel. Accordingly, it should be possible to estimate parameters of the spiral arms like pattern speed and relative spiral phase. However it may be necessary to invoke both dynamical and accretion models to explain velocity distributions in  the solar neighborhood.

Metallicity distributions can be of great help for understanding origins of these streams. A uniform metallicity would be expected for a cluster origin, so a large range of metallicity scatter or multiple populations can rule out that possibility. Only by detailed abundance analysis can a stellar streams's origin be determined with any degree of certainty.

Soon, by combining the Gaia TGAS and LAMOST DR3 data, we will be able to confidently detect fine structures in the solar vicinity. For now, using the wavelet technique, sixteen significant overdensities were detected with high confidence from a sample of 27290 stars. Eleven of them correspond to the previously known streams detected in the RAVE project \citep{ant12}. The identity of peak 15 is in doubt. The kinematics and metallicities of these streams are consistent with previous works. Four new stream candidates named New1, New2, New3, New4 were found. New1 (peak 9) and New2 (peak 12) most likely are the results of the dynamical effect of the Galaxy; while the origins of NEW3 (peak 13) and NEW4 (peak 14) are uncertain. Detailed abundance analyses are needed to better understand the origins of these streams. Unveiling the origins of NEW3 and NEW4 may provide observational constraints on gravitational potential models of the Milky Way.

\acknowledgments
We thank Xue Xiangxiang and Li Guangwei for their constructive suggestions and discussion. This study is supported by the National Natural Science Foundation of China under grant No. 11390371, 11233004, U1431106, 11573035, 11625313, the National Key Basic Research Program of China (973 program) 2014CB845701 and 2014CB845703 and support from the US National Science Foundation (AST-1358787) to Embry-Riddle Aeronautical University is acknowledged. Guoshoujing Telescope (the Large Sky Area Multi-Object Fiber Spectroscopic Telescope LAMOST) is a National Major Scientific Project built by the Chinese Academy of Sciences. Funding for the project has been provided by the National Development and Reform Commission. LAMOST is operated and managed by the National Astronomical Observatories, Chinese Academy of Sciences.

%\email{aastex-help@aas.org}.

\clearpage
\begin{figure}
\begin{minipage}{0.8\linewidth}
  \centerline{\includegraphics[scale=0.6]{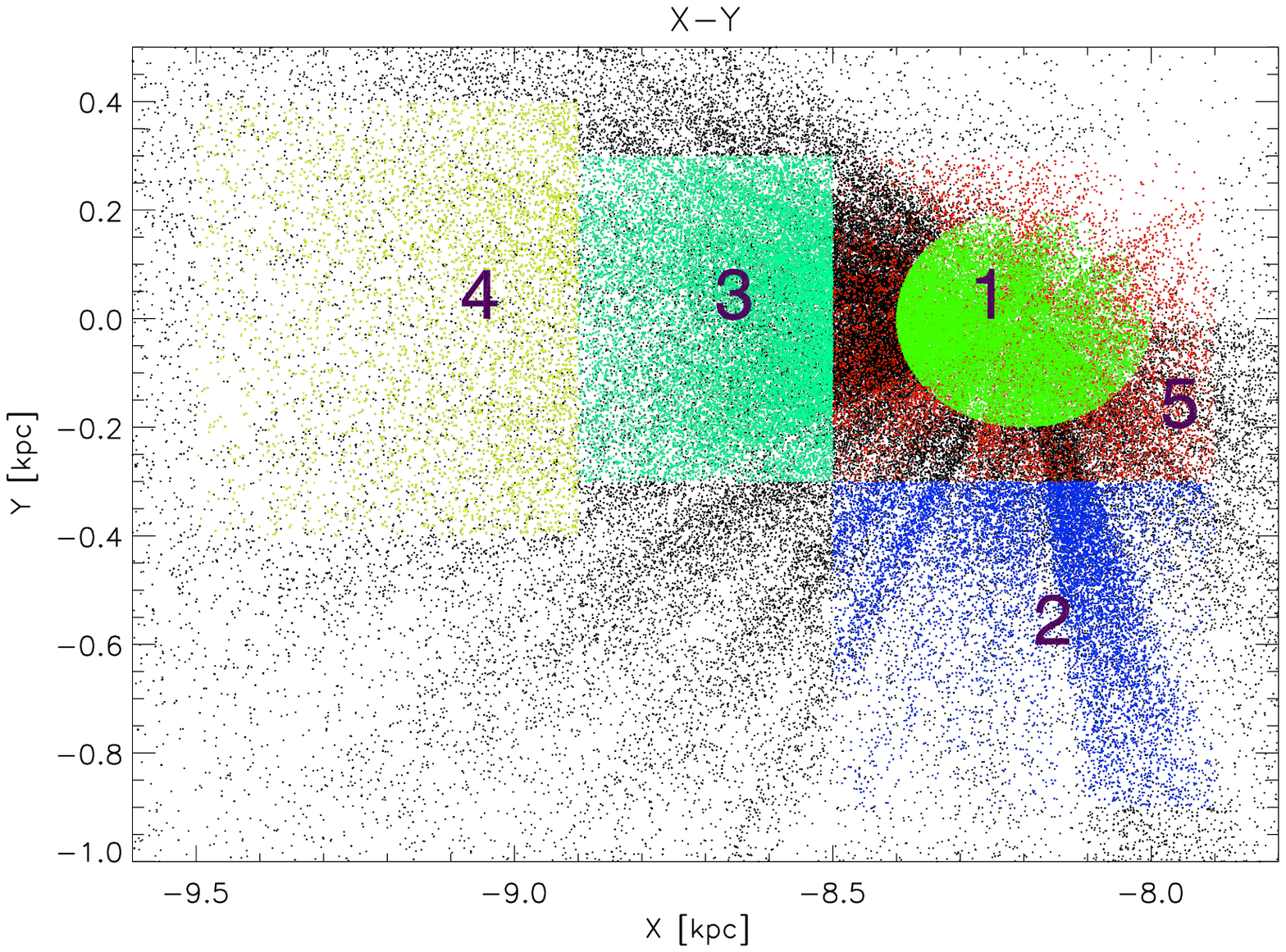}}
  %\centerline{xy}
\end{minipage}
\vfill
\begin{minipage}{0.8\linewidth}
  \centerline{\includegraphics[scale=0.6]{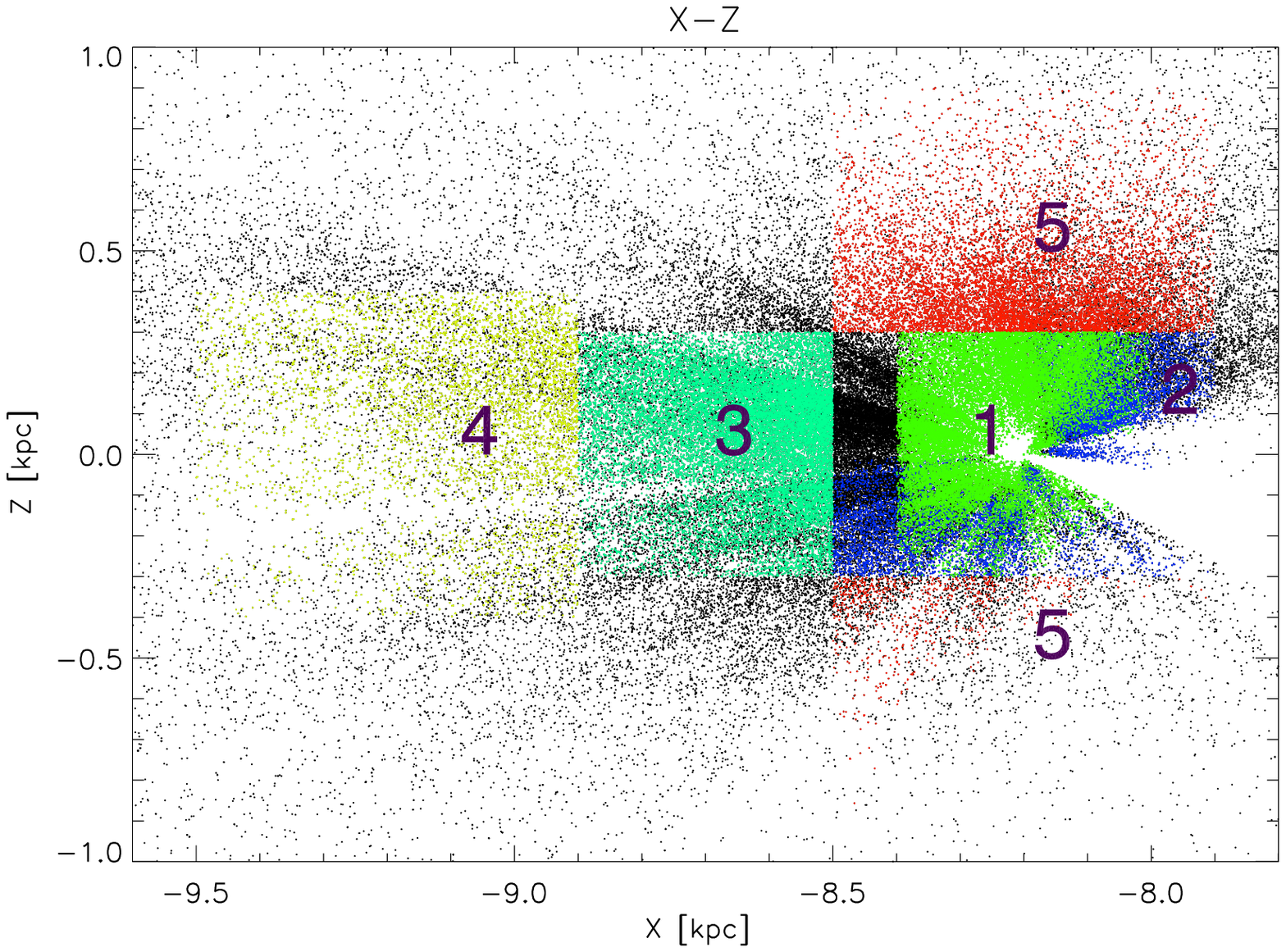}}
%  \centerline{xz}
\end{minipage}
\caption{X, Y, Z components of stars in the total sample. Colors corresponding to each region are 1-lime, 2-blue, 3-aquamarine, 4-yellow, 5-red. See text for details.\label{Figure 1}}
\end{figure}

\begin{figure}
\epsscale{.90}
\plotone{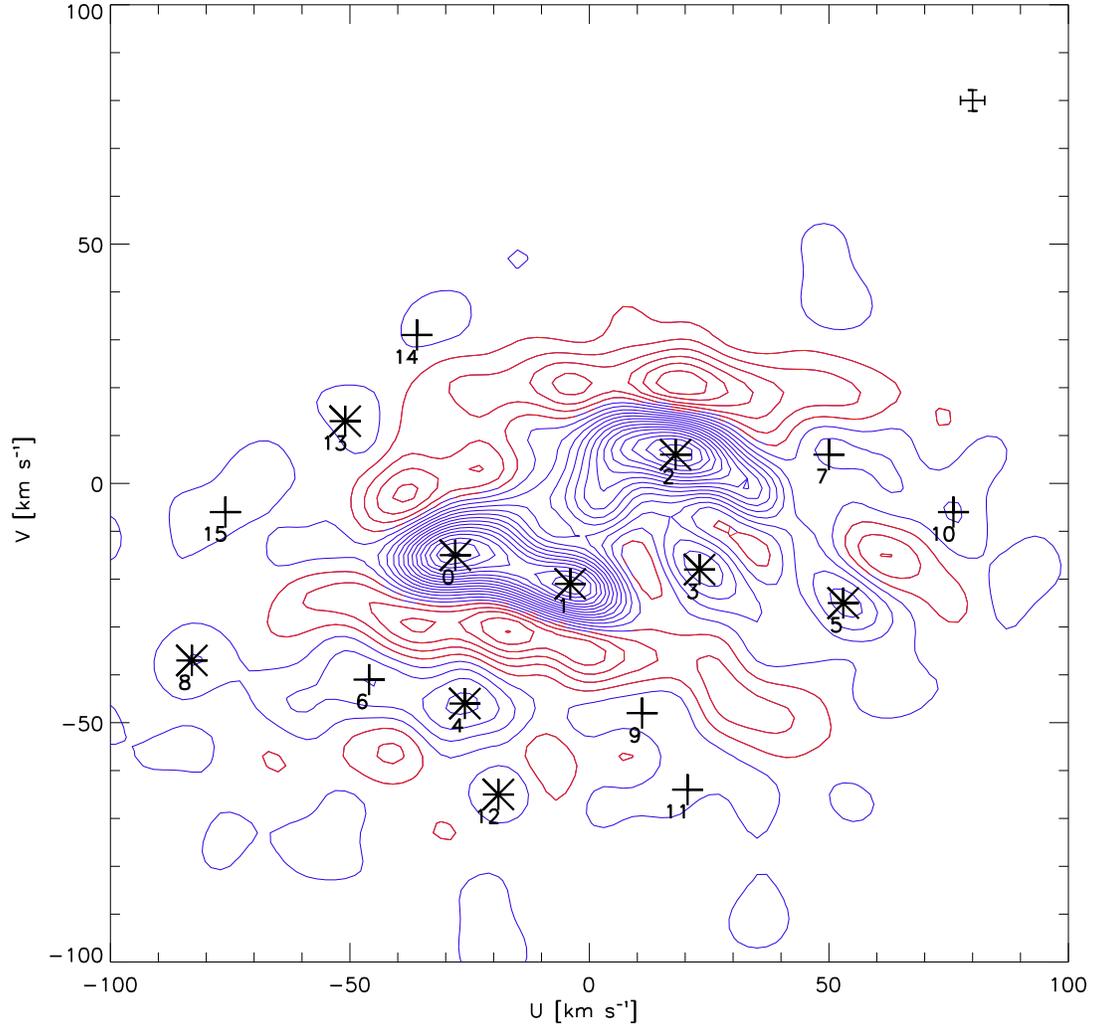}
\caption{Wavelet transform of the UV velocity distribution of the LS with a scale of $\thicksim $6 km $\rm s^{-1}$. The typical error bar for U and V is in the upper-right corner. The mean errors of U and V are 5.05 and 4.37 km $\rm s^{-1}$ respectively. See text for details \label{Figure 2}}
\end{figure}

\begin{figure}
\epsscale{.80}
\plotone{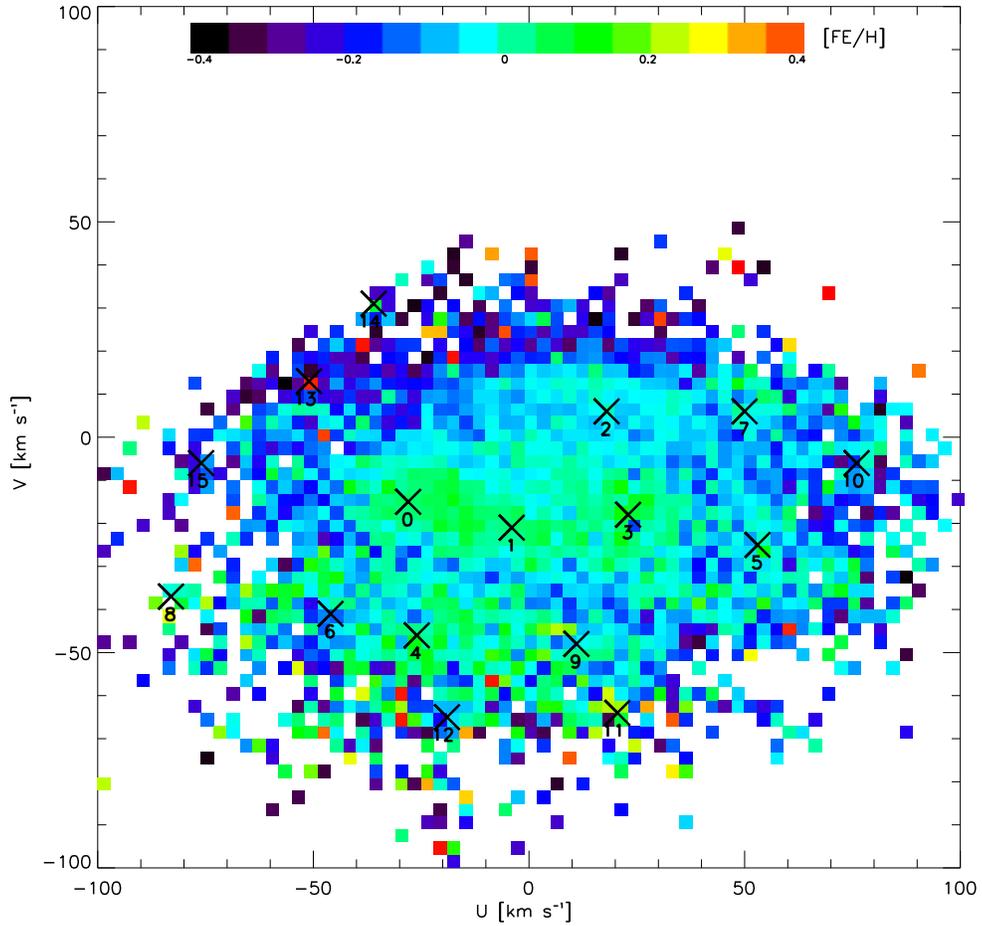}
\caption{The metallicity distribution of the LS described in the text. The color scale indicates the mean metallicity of each bin. The blank regions indicate regions containing fewer than 4 stars in a bin. The typical uncertainty in metallicity is 0.125. See text for details.\label{Figure 3}}
\end{figure}

\begin{figure}
\begin{minipage}{0.45\linewidth}
  \centerline{\includegraphics[scale=0.44]{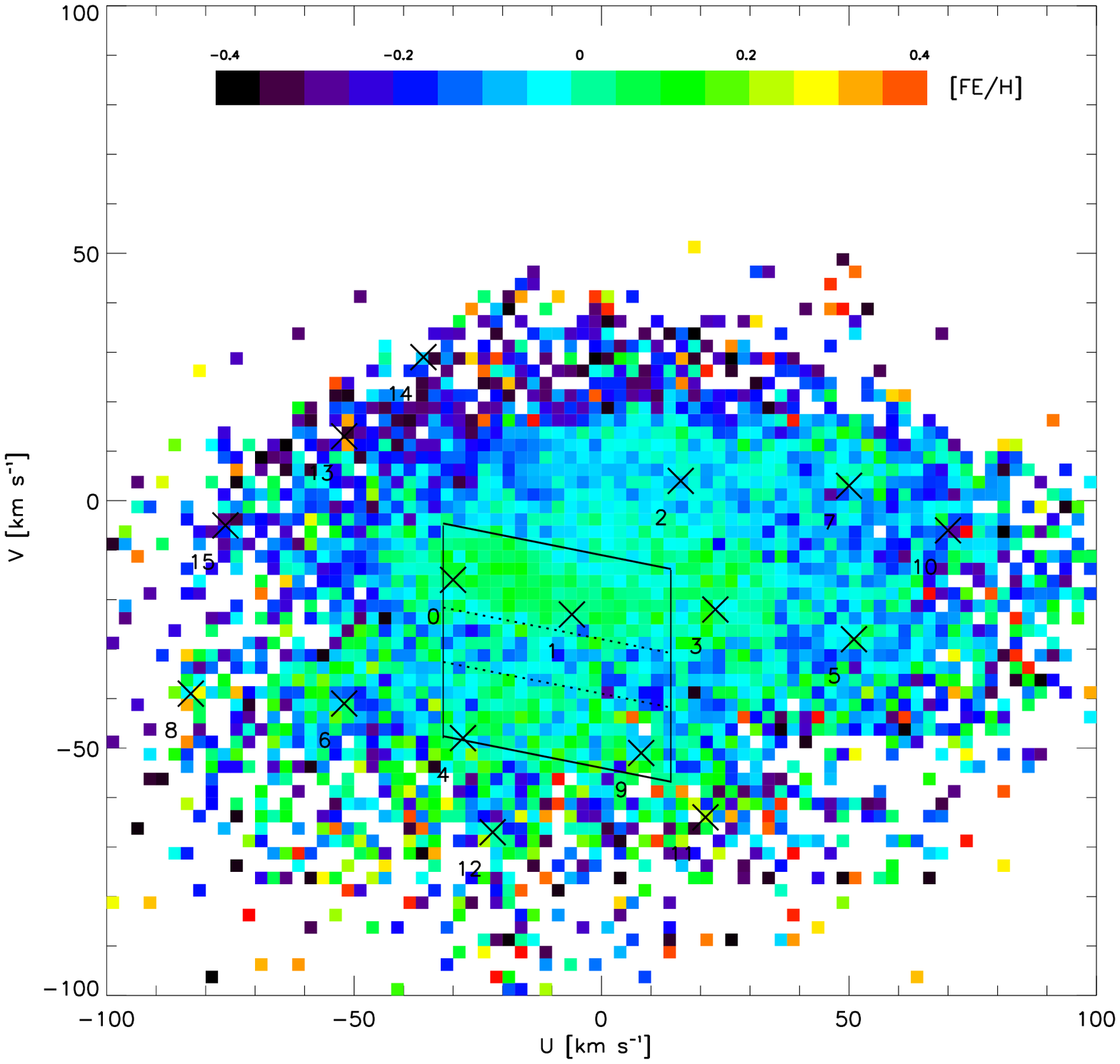}}
\end{minipage}
\hfill
\begin{minipage}{0.45\linewidth}
  \centerline{\includegraphics[scale=0.425]{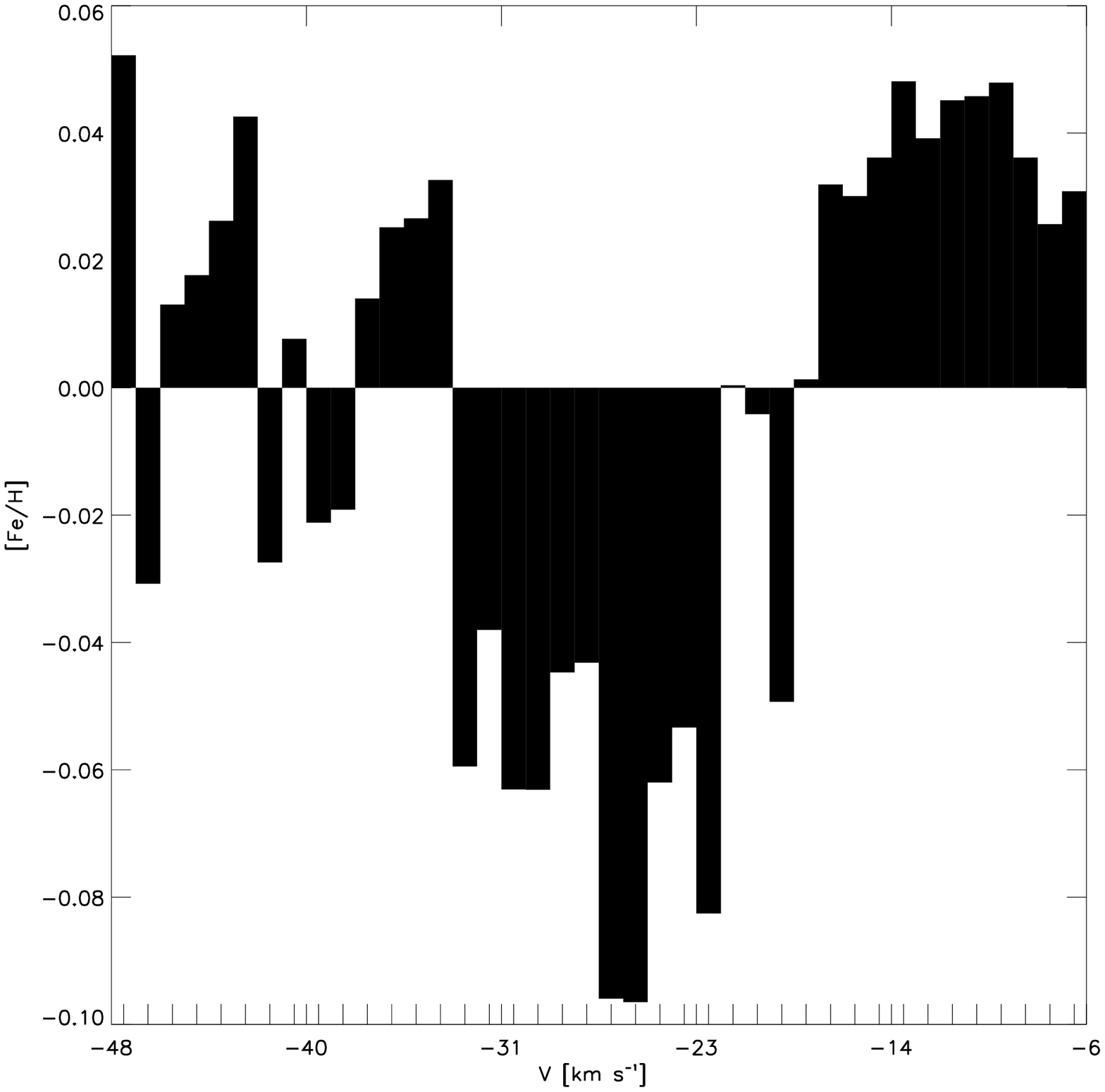}}
\end{minipage}
\caption{The left panel shows the window we picked in the metallicity distribution of the LS. And the right panel shows the mean metallicity of each bin distributes along with the vertical direction. The bin size is 1 km $\rm s^{-1}$. See text for details.\label{Figure 4}}
\end{figure}

\begin{figure}
\begin{minipage}{0.48\linewidth}
  \centerline{\includegraphics[scale=0.4]{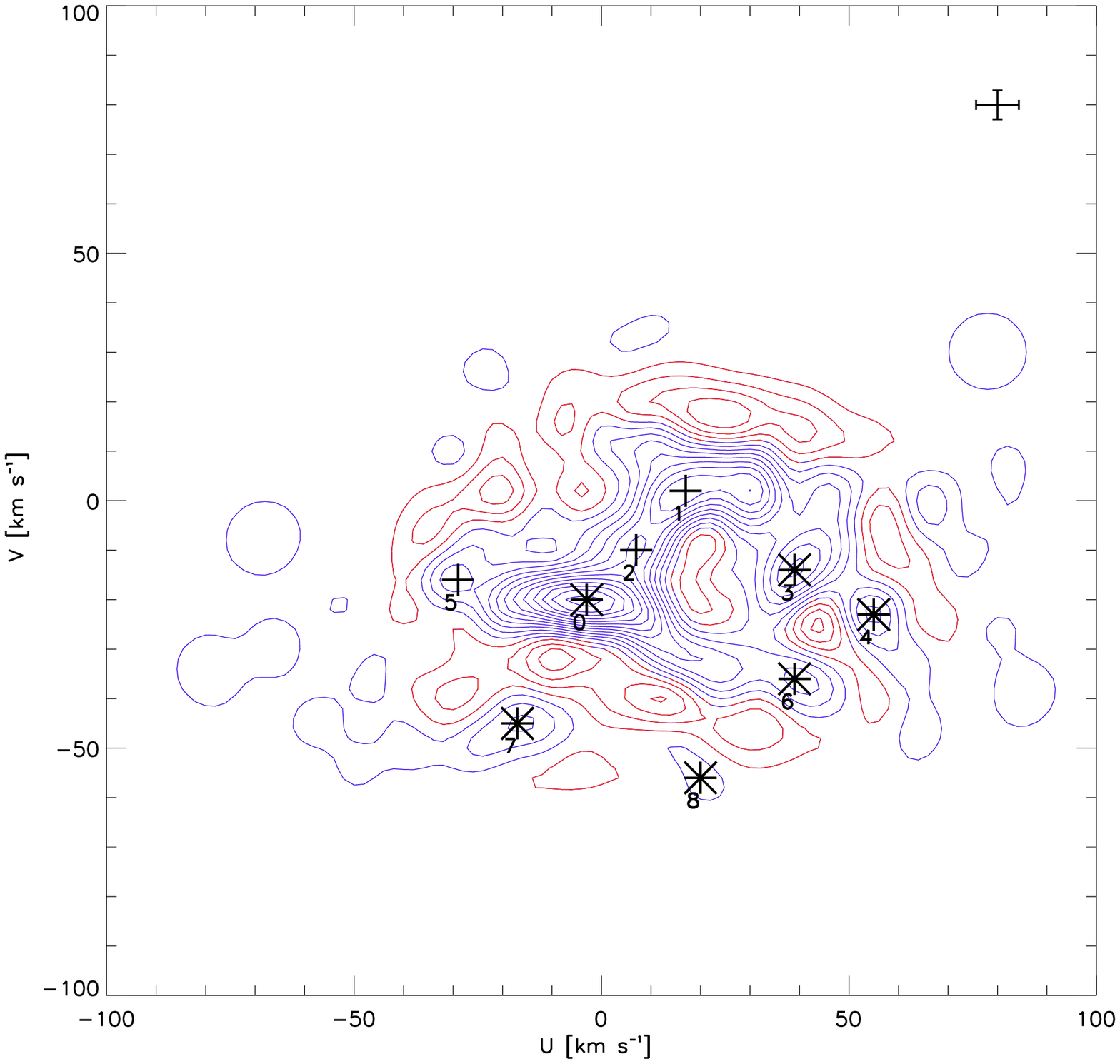}}
  \centerline{(a) region 2}
\end{minipage}
\hfill
\begin{minipage}{.48\linewidth}
  \centerline{\includegraphics[scale=0.4]{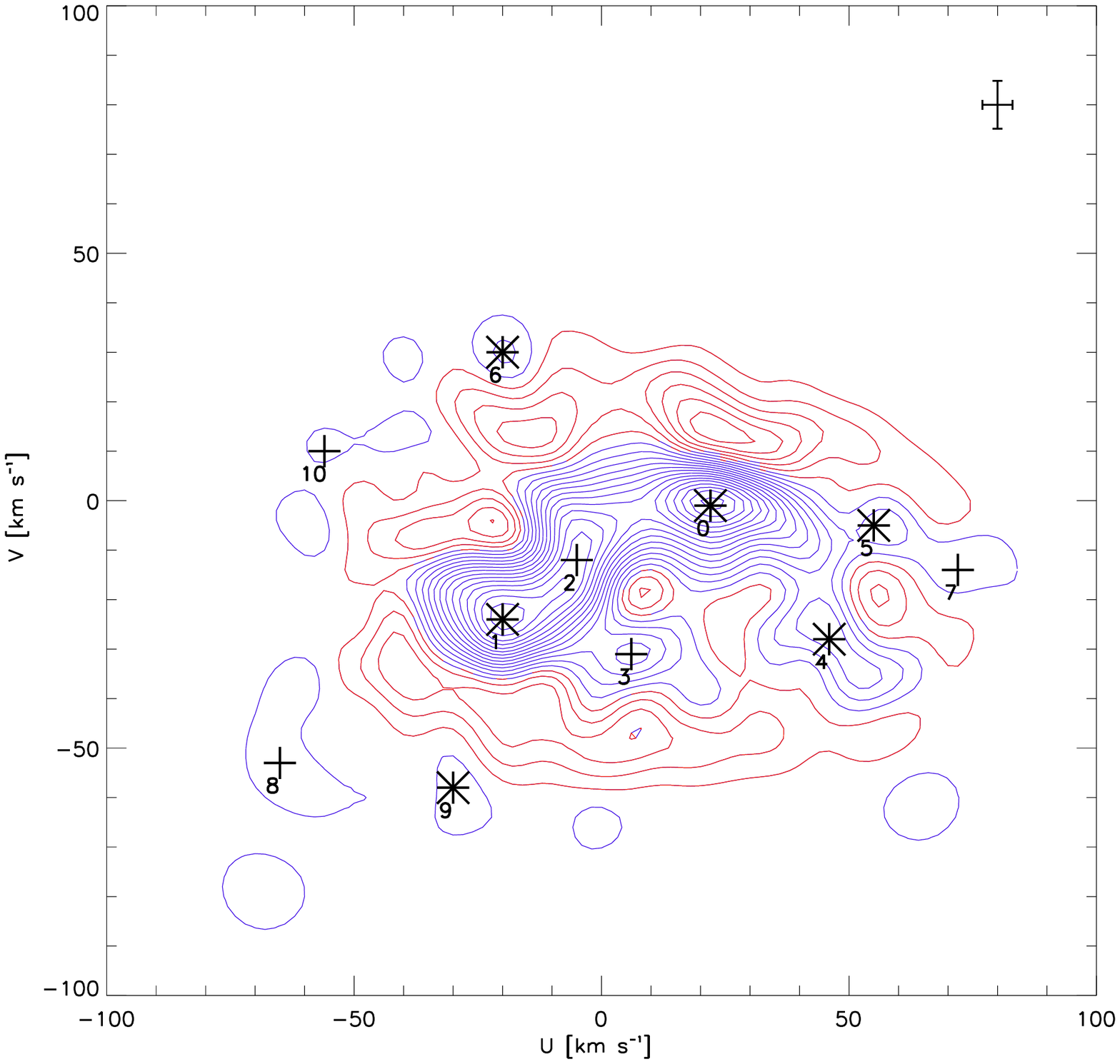}}
  \centerline{(b) region 3}
\end{minipage}
\vfill
\begin{minipage}{0.48\linewidth}
  \centerline{\includegraphics[scale=0.4]{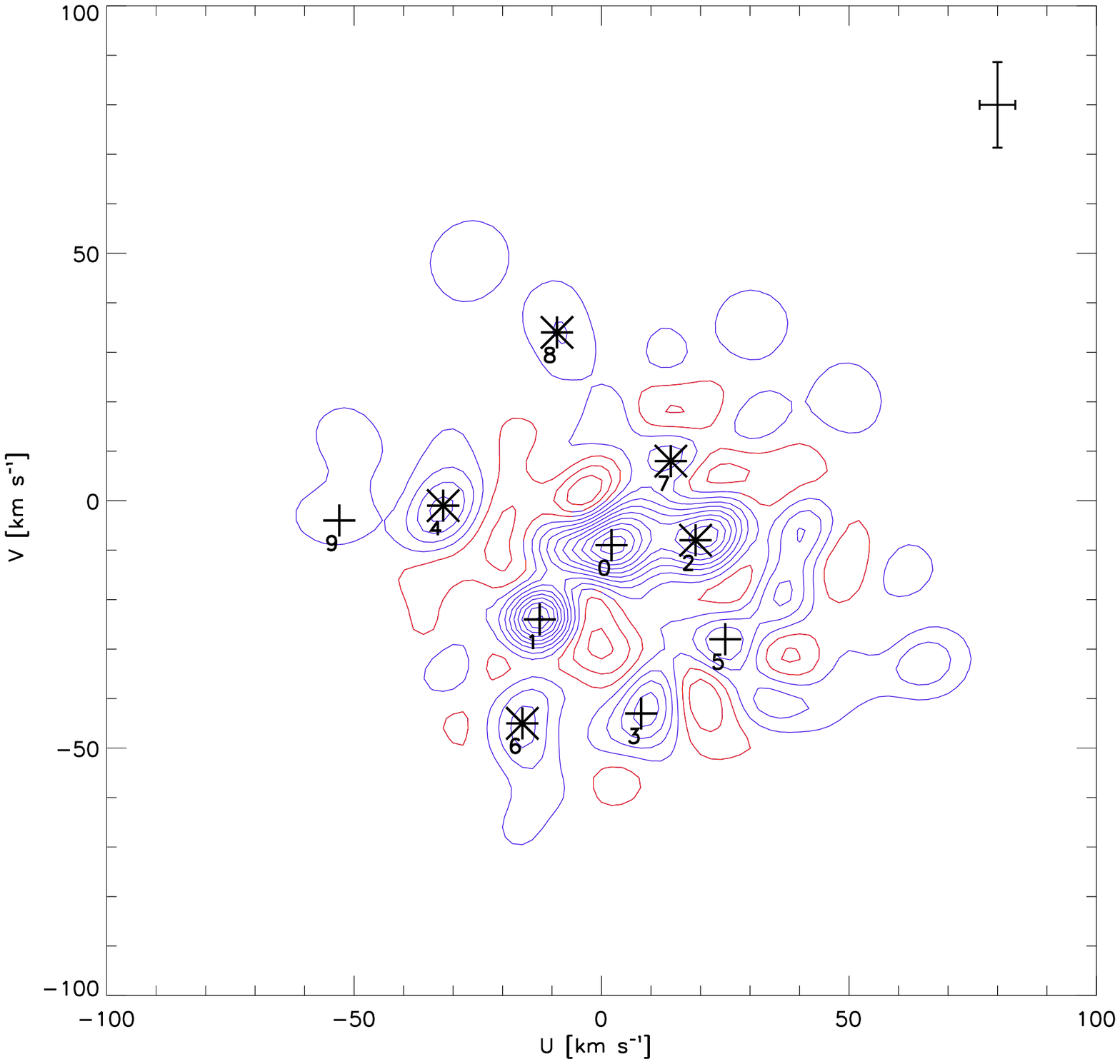}}
  \centerline{(c) region 4}
\end{minipage}
\hfill
\begin{minipage}{0.48\linewidth}
  \centerline{\includegraphics[scale=0.4]{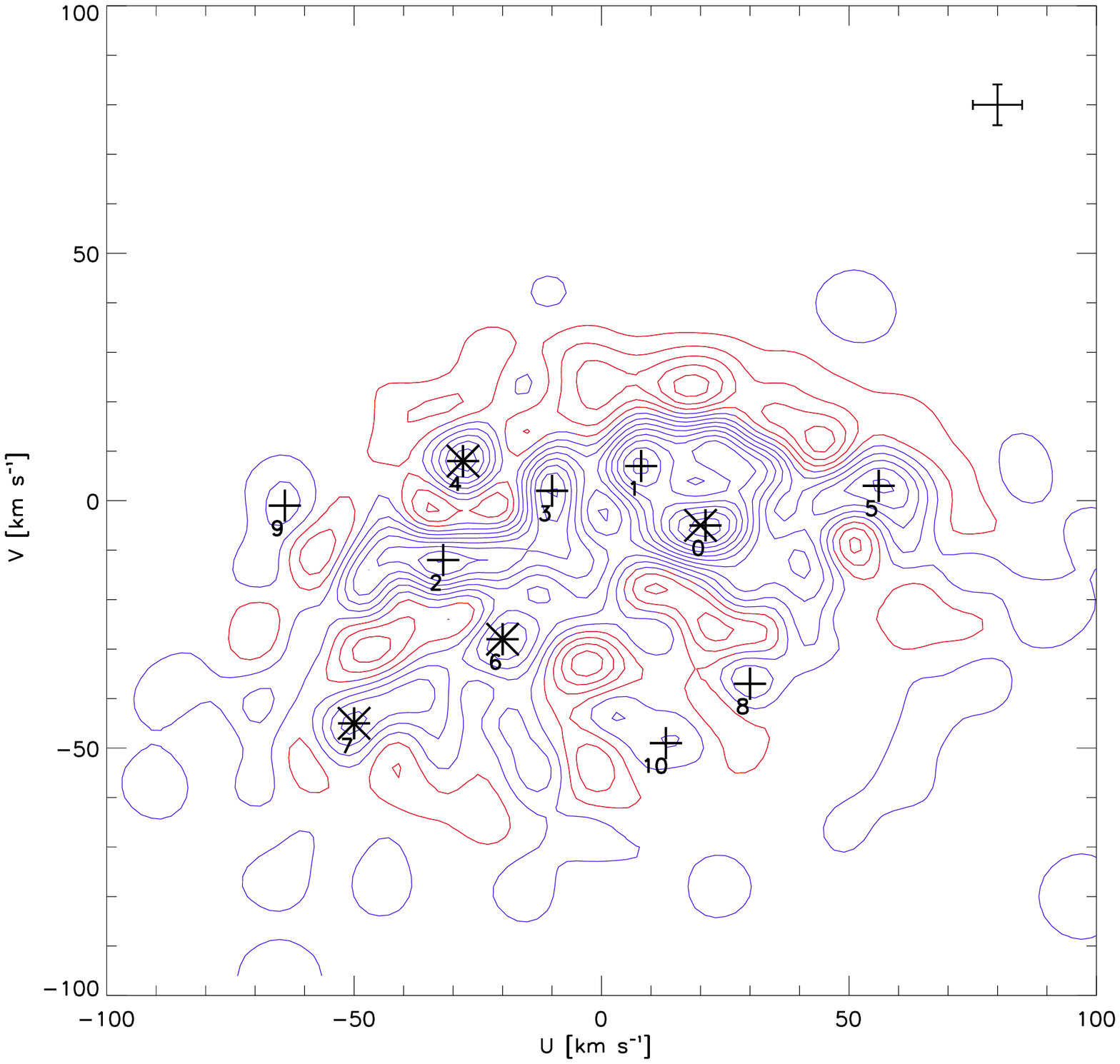}}
  \centerline{(d) region 5}
\end{minipage}
\caption{Wavelet transform of velocity distribution of subsamples beyond the LS. The typical error bar for U and V is in the upper-right corner of each panel. The mean errors of U and V in (a) are  8.65 and 5.85 km $\rm s^{-1}$ respectively. The mean errors of U and V in (b) are 6.06 and 9.69 km $\rm s^{-1}$ respectively. The mean errors of U and V in (c) are 7.24 and 17.31 km $\rm s^{-1}$ respectively. The mean errors of U and V in (d) are 9.96 and 8.24 km $\rm s^{-1}$ respectively. See text for details.\label{Figure 5}}
\end{figure}

\begin{deluxetable}{cccccccc}
\tablecolumns{8}
\tablewidth{0pc}
\tablecaption{ Kinematic groups detected in the local sample. The columns give,respectively: the WT coefficient order, name of the group in published literature (if any), U velocity and V velocity of peaks, approximate number of stars, and the mean and variance of metallicity from Gaussian fitting. The last column shows the equivalent groups in the RAVE survey \citep{ant12}. Velocities are in km $\rm s^{-1}$.}

\tablehead{
\colhead{}    &  \multicolumn{6}{c}{TGAS \& LAMOST} & \colhead{RAVE}  \\
\colhead{No.} & \colhead{ Name}   & \colhead{U }    & \colhead{V } &
\colhead{N}   & \colhead{[Fe/H]}  & \colhead{$\rm \sigma_{[Fe/H]}$}   & \colhead{ No.}
}
\startdata

        0 & Hyades       & -30 & -16 & 1311  & 0.060  & 0.207 & 2 \\
        1 & Pleiades     & -6  & -23 & 1261 & 0.048  & 0.186 & 4 \\
        2 & Sirius       & 16  &  4  & 1409 & -0.033 & 0.157 & 3 \\
        3 & Wolf 630     & 23  & -22 & 927  & 0.064  & 0.224 & 5 \\
        4 & Hercules \uppercase\expandafter{\romannumeral2}  & -28 &  -48 & 396  & 0.022 & 0.259 & 8 \\
        5 & Dehnen 98    & 51  & -28 & 491  & -0.018 & 0.196 & 7 \\
        6 & Hercules \uppercase\expandafter{\romannumeral1}  & -52 & -41 & 276  & -0.044 & 0.250 & 6 \\
        7 & $\gamma$ Leo  &  50  &  3  & 395  & -0.073 & 0.221 & 9 \\

        8 & $\epsilon$ Ind & -83 & -39 &  71  & -0.038  & 0.237 & 10 \\
        9 & NEW1          &  8  &  -51 & 296  & -0.005  & 0.244 & \\
        10 & $\gamma$ Leo  & 70 &  -6  & 232  & -0.106 & 0.193 & 11 \\
        11 & HR1614        & 21  & -64 & 65  &  0.061 & 0.403 &   \\
        12 & NEW2          & -22 & -67 & 143  & -0.104 & 0.334 &  \\
        13 & NEW3          & -52 & 13  & 141  & -0.218 & 0.238 & \\
        14 & NEW4          & -36 & 29  &  72  & -0.247 & 0.197 & \\
        15 &               & -76 & -5  & 93   & -0.207 & 0.191 & \\

\enddata
\end{deluxetable}

\end{document}